\documentclass[11pt, letterpaper]{article}

\usepackage[utf8]{inputenc}
\usepackage[margin=1in]{geometry}
\usepackage{booktabs}

\usepackage[labelfont=bf, skip=5pt, font=small]{caption}

\usepackage[]{graphicx}

\usepackage{natbib}

\usepackage{amsmath,amssymb}

\usepackage{textcomp} 

\begin{document}

\begin{center}
  \huge{Multiple power laws and scaling relation \\ in exploratory locomotion of the snail\\ \textit{Tegula nigerrima}}
  \par

\large{Katsushi Kagaya}\footnote{
  National University Corporation Hokkaido Higher Education and Research System, Kitami Institute of Technology \\
  165 Koen-cho, Kitami, Hokkaido 090-8507, Japan\\
  E-mail: kkagaya@mail.kitami-it.ac.jp\\
}, 
  \large{Tomoyuki Nakano}\footnote{
  Seto Marine Biological Laboratory, Field Science Education and Research Center, Kyoto University\\
  459 Shirahama, Nishimuro, Wakayama 649-2211, Japan\\
  }, 
  \large{Ryo Nakayama}\footnote{
  Fishries Research Institute, Aomori Prefectural Industrial Technology Research Center\\
  10 Tsukidomari Moura, Hiranai, Higashi-Tsugaru, Aomori 039-3381, Japan\\
  }

\end{center}

\begin{abstract}
  \noindent 
  One of goals in soft robotics is to achive spontaneous behavior like real organisms. 
  To gain a clue to achieve this, we examined the long (16-hour) spontaneous exploratory locomotion of snails. 
  The active forager snail, \textit{Tegula nigerrima}, from an intertidal rocky shore was selected to test the general hypothesis that nervous systems are inherently near a critical state, which is self-organized to drive spontaneous animal behavior. 
  This hypothesis, known as the critical brain hypothesis, was originally proposed for vertebrate species, but it might be applicable to other invertebrate species as well.
  We first investigated the power spectra of the speed of locomotion of the snails ($N=39$). 
  The spectra showed $1/{f^\alpha}$ fluctuation, which is one of the signatures of self-organized criticality. 
  The $\alpha$ was estimated to be about 0.9. 
  We further examined whether the spatial and temporal quantities show multiple power-laws and scaling relations, which are rigorous criteria of criticality. 
  Although the satisfaction of these criteria is limited to a truncated region and provides limited evidence to demonstrate the aspect of self-organization, the multiple power-laws and the scaling relations were overall satisfied. 
  Therefore, these results additionally support the generality of the critical brain hypothesis.

\end{abstract}

\pagebreak

\section{Introduction}
Biological organisms have long inspired engineers to achieve artificial robotic agents (\cite{pfeifer2007self,suzumori2022overview}). 
Spontaneity is a difficult property to implement because it seems far from stability or balance by definition. 
Additionally, it appears difficult to control or design externally. 
However, real organisms seem self-organized and behave spontaneously without such external inputs. 
This property evolved and developed in nature, but we do not know what the physical mechanism is. 
In biology, spontaneous exploratory movements have long been extensively studied, and it is well known that the movement patterns can be well described by a power-law distribution, and such pattern is called  L\'{e}vy walk (\cite{shlesinger1993strange,sims2008scaling,sims2014hierarchical,reynolds2018current}).
One scenario of the generative process is a mechanism of nonlinear dynamical systems showing critical dynamics (\cite{maye2007order,abe2020functional,o2022critical}). 
Indeed, the mechanism was proposed and demonstrated (\cite{abe2020functional}). 
The functional advantages of the spontaneous movement pattern with power-law characteristics in their temporal and spatial quantities were demonstrated in a model of nonlinear dynamical systems (\cite{abe2020functional}). 
The term L\'{e}vy walk alone does not imply its generative mechanism. 
It is simply realized as a random sample from a power-law distribution.
In contrast, the fruit fly research (\cite{abe2020functional}) provided us with an alternative view from nonlinear dynamical systems. 
The model includes an order parameter that needs to be tuned to provide critical dynamics. 
When a system is at the critical state, any physical quantities follow power-laws.
However, how is the order parameter tuned?

Real organisms may not have such parameters. 
Real organisms have individual differences that emerge via evolutionary, developmental, and physiological processes. 
These differences can arise even when the genetic backgrounds are initialized in the same state. 
Despite these differences, the critical state seems to be achieved. 
Bak and his colleagues conjectured that this state is achieved in living systems without some order parameters (\cite{bak1987self,bak1988self,bak1989self}). 
They termed this naturally evolved state as self-organized criticality.

In this study, we focus on an active, representative soft-bodied animal, the snail. 
Specifically, the snail \textit{Tegula nigerrima} is studied. 
It is an active mollusc in the intertidal rocky shore. 
Its spontaneous exploratory behavior is considered one factor for maintaining the diversity of the environment. 
For example, this spontaneous activity is exploited by smaller and slower limpets (\cite{nakayama2020seasonal}). 
The snail moves around the complex environment, but the dynamical properties behind its locomotion are unclear.


We hypothesize that the origin of the power law is the self-organized criticality in the nervous systems. 
To examine this hypothesis, we first investigate whether the spontaneous locomotor behavior shows multiple power laws: power spectrum, speed (step length of one frame), distance, and duration of walking bouts. 
Specifically, the last two quantities, distance and duration, can provide a spatial and temporal scaling relation (\cite{jensen1998self,sethna2001crackling}). 
We test whether this relationship is satisfied. 
Through the scaling study, we examine whether different individuals show some common properties. 
This is important because if different individuals with varying natural genetic and developmental backgrounds show some common properties, it suggests that the property is self-organized. 
Additionally, we prepared experimental conditions close to a real environment, but controlled to be constant, to examine spontaneous behavior.

\section{Materials and Methods}

\subsection{Animal}
We collected the snails \textit{Tegula nigerrima} on intertidal rocky shores near Seto Marine Biological Laboratory (SMBL), Shirahama, Wakayama Prefecture, Japan. 
The snails ($N=39$) were from a lower intertidal rocky boulder shore (Kitahama site;33\textdegree41\textquotesingle40.7\textquotesingle\textquotesingle N, 135\textdegree20\textquotesingle12.9\textquotesingle\textquotesingle E).

\subsection{Experiment}
A container with dimensions of 10 cm in width, 7.6 cm in depth, and 10 cm in height was prepared. 
Three holes of 4 mm for one side plane were made at a height of 3.5 cm from which seawater could flow out. 
The container was used to hold running seawater for the purpose of recording the free walking behavior of a snail for 16 hours using a video camera. 
The snail's movements were recorded at 0.5 fps using a GoPro HERO7. 
The recorded video was then analyzed using UMA tracker (\cite{yamanaka2018umatracker}) to track and document the snail's motion as planar motion. 
The distinction between vertical motion and stationary state in the plane could not be made.

For nineteen cases, the crab \textit{Leptodius affinis} were induced but no direct contact was possible between them. 
They interacted chemically through seawater only. 
Four tiny limpets \textit{Lottia tenuisculpta} were placed on platforms with a height of 5 mm and a radius of 14 mm. 
This setup approximates intertidal conditions while maintaining as constant a state as possible. 
\textit{Lottia tenuisculpta} exhibited hitchhiking behavior by riding on the gastropods but rarely interfered with the movement behavior of the snails. 
In fact, no significant differences were found in the velocity time series or their power spectra depending on the presence of the crabs.

\subsection{Data Analysis}
Power spectrum analysis, power-law fitting, and scaling relation data analysis were conducted using the statistical analysis environment R 4.1.2. 
The power-law fitting results are shown with appropriately chosen bin widths for the histograms. 
The impact of the sampling rate and the data length limitation on the time series is considered to be smaller than in any previous reports. 
This is due to the high sampling rate, long data length, and large sample size.

\section{Results}

\begin{figure*}[h]
  \centering
  \includegraphics[width=15cm]{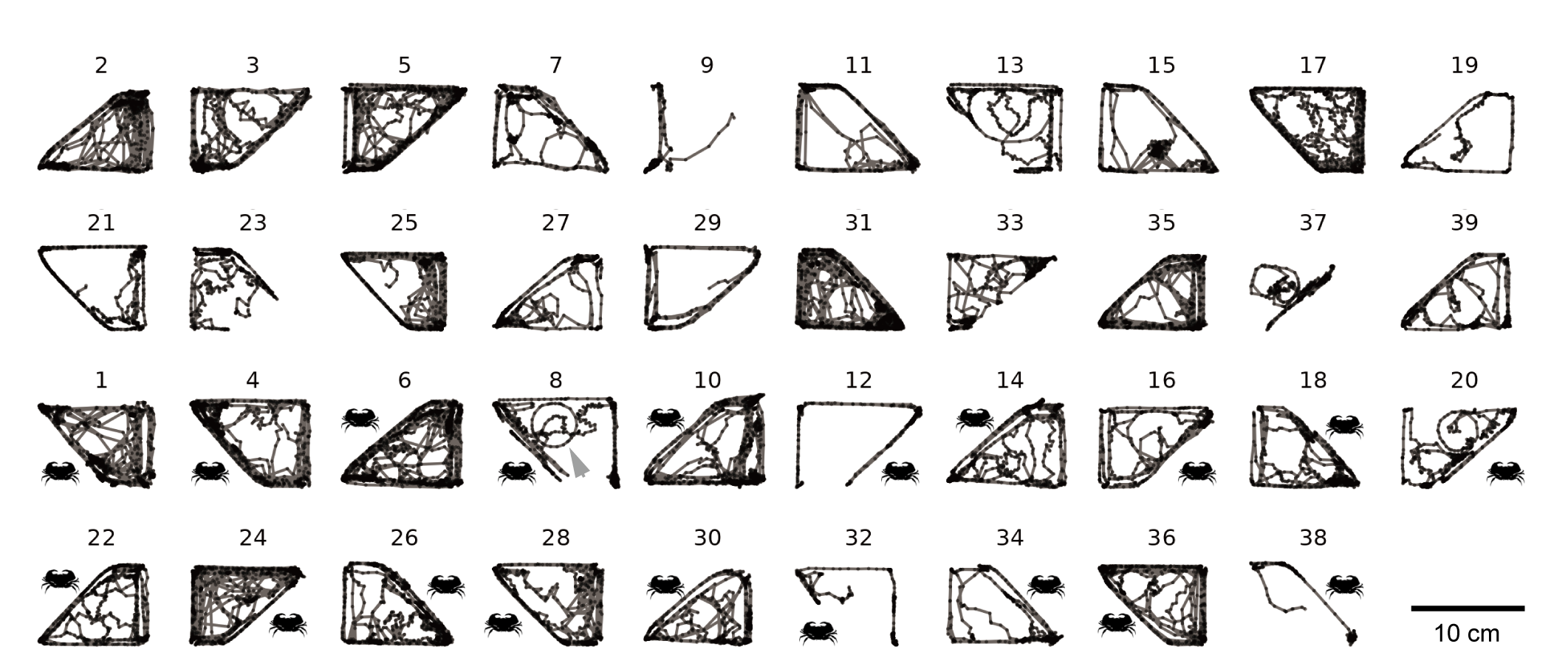}
  \vspace{-5pt}
  \caption{Tracked paths of the snails. Carnivorous crabs were present in a half of conditions. They were only chemically connected to the snails via holes on partitions.}
  \label{fig_trajectory}
\end{figure*}

\begin{figure*}[h]
  \centering
  \includegraphics[width=15cm]{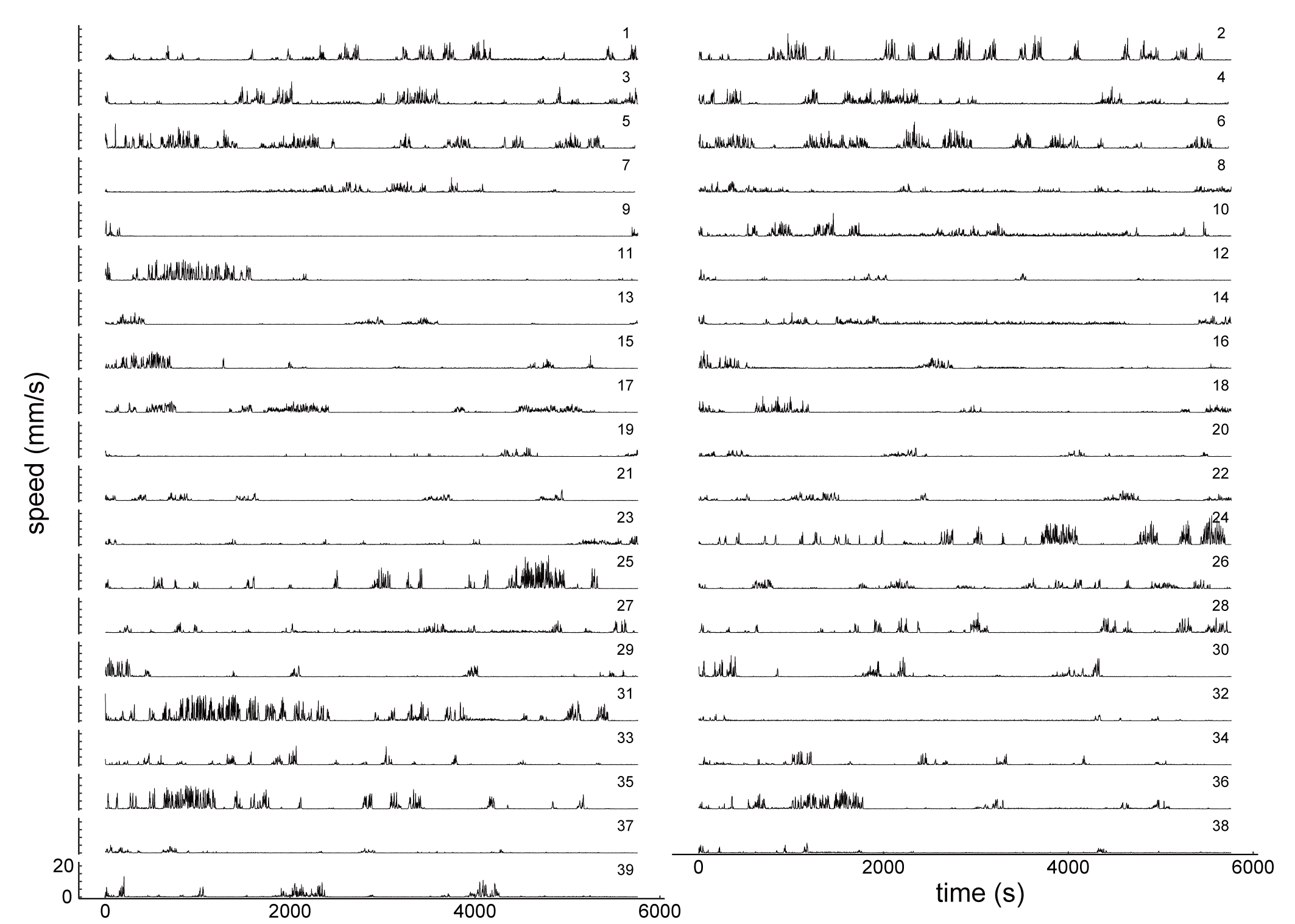}
  \vspace{-5pt}
  \caption{Change of spontaneous locomotor speed of the snails.}
  \label{fig_speed}
\end{figure*}

Each individual moved around in the container with its own distinct behavior (Fig.\ref{fig_trajectory}).
No effects of the presence of crabs were found, and it appears that individual differences contributed more to the variation in movement patterns. 
Similarly, significant variations due to individual differences were observed in the speed time series as well (Fig.\ref{fig_speed}).

\begin{figure*}[t]
  \centering
  \includegraphics[width=15cm]{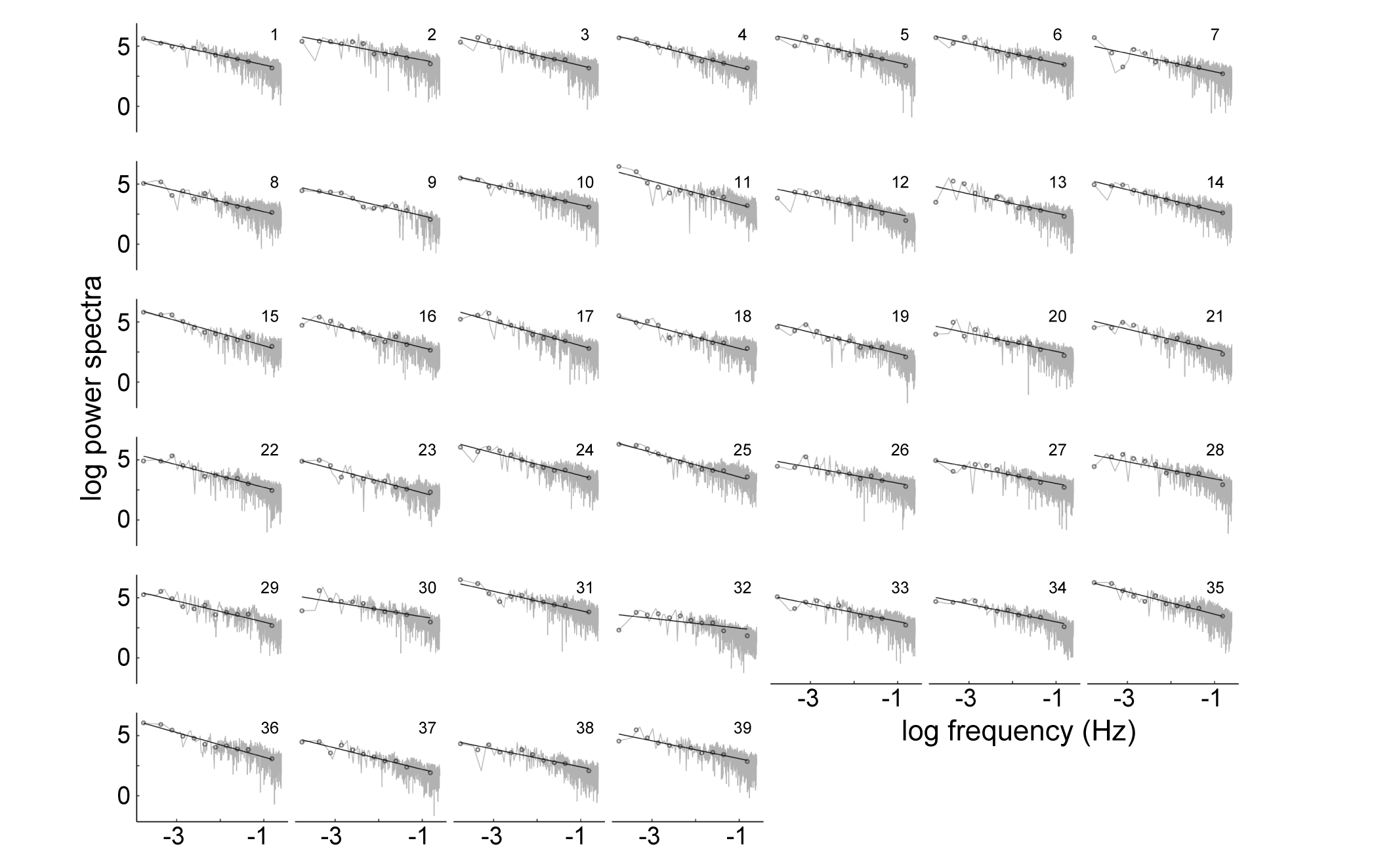}
  \vspace{-5pt}
  \caption{Power spectra of spontaneous locomotor speeds of the snails.}
  \label{fig_power_spectra}
\end{figure*}

\begin{figure}[t]
  \centering
  \includegraphics[width=7.5cm]{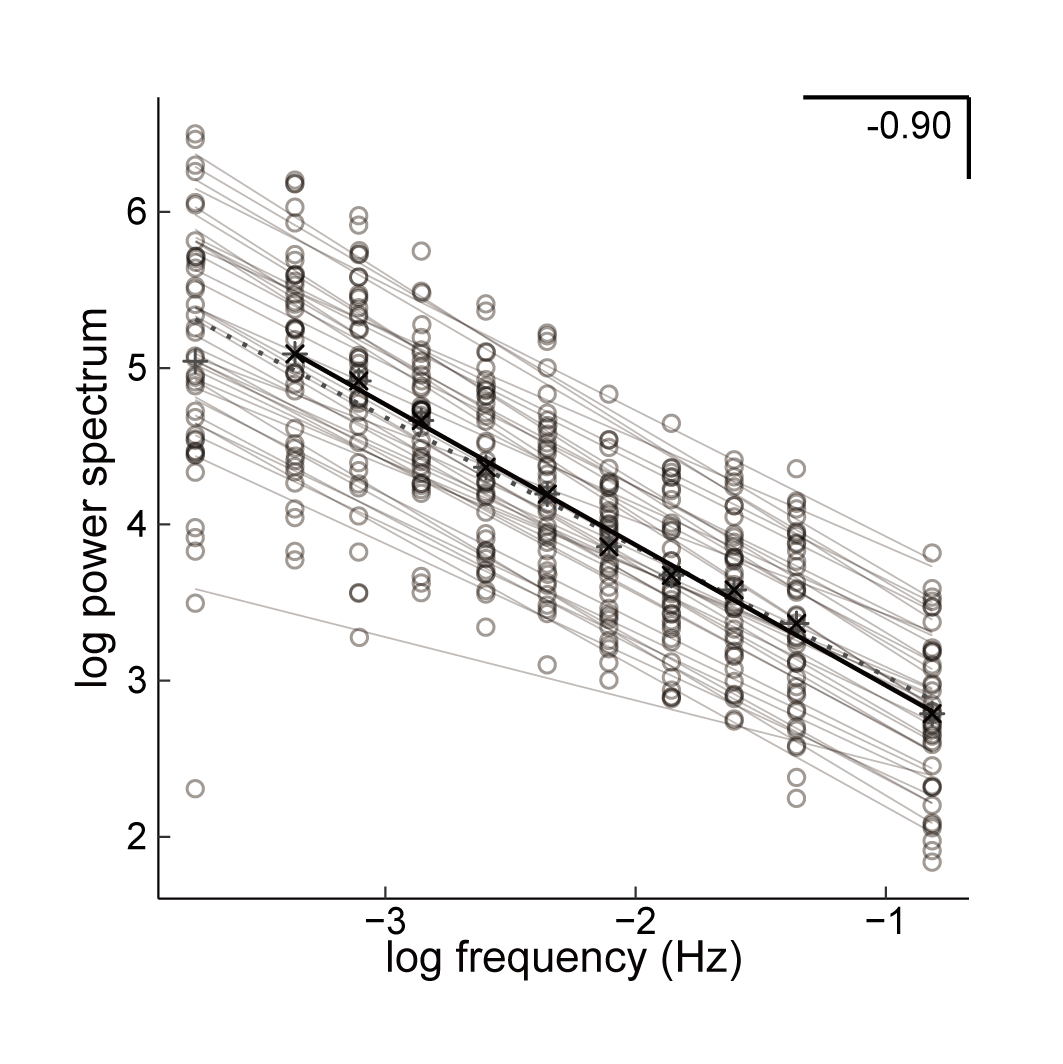}
  \vspace{-5pt}
  \caption{Estimation of an exponent of the power spectrum.}
  \label{fig_ps_sum}
\end{figure}

Despite the individual differences in the trajectories and speed time series, the power spectra were very similar (Figs.\ref{fig_power_spectra}-\ref{fig_ps_sum}).
This exponent was calculated as about 0.9, thus a little smaller than unity (Fig.\ref{fig_ps_sum}). 
Thus, the power spectra overall showed $1/f^{0.9}$.

\begin{figure}[t]
  \centering
  \includegraphics[width=7.5cm]{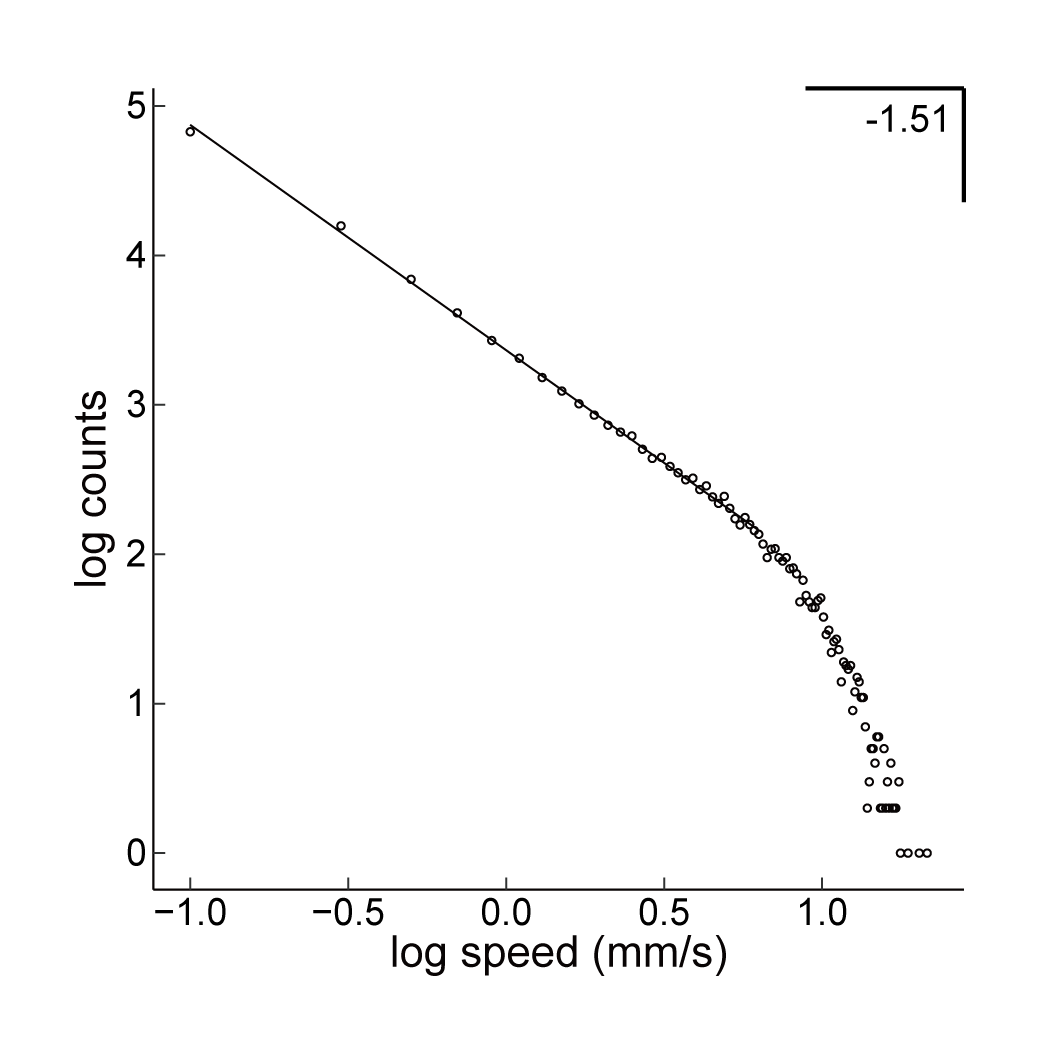}
  \vspace{-5pt}
  \caption{Distribution of speeds (step lengths per a frame).}
  \label{fig_speed_dist}
\end{figure}

We pooled the all time series and conducted power-law fitting to the distribution of speeds (Fig.\ref{fig_speed_dist}). 
The exponent was estimated as 1.51, thus it is a typical L\'{e}vy walk movement pattern considered optimal for foraging behavior when resources are patchy.

\begin{figure*}[t]
  \centering
  \includegraphics[width=15cm]{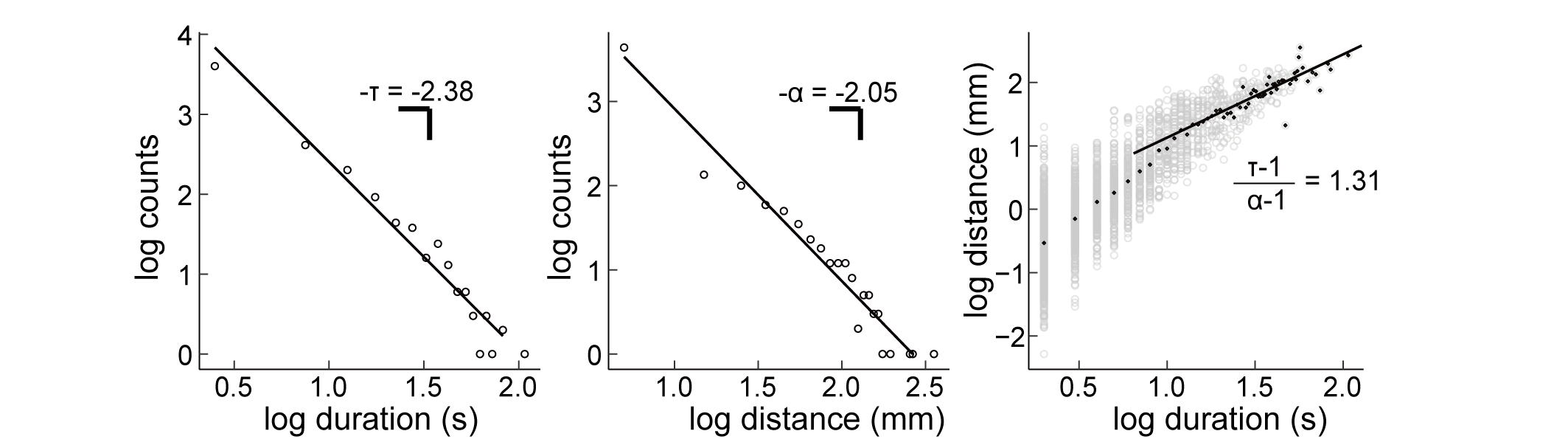}
  \vspace{-5pt}
  \caption{Crackling noise relation of the exponents. 
    The slope of the line in the right was calculated from  $\tau$ and $\alpha$. 
    The exponents represent temporal and spatial aspects. 
    The closed circles are median values for each duration and the line with the slope of 1.31 match the circles. 
    We consider that this match is an evidence of criticality.
  }
  \label{fig_scaling_relation}
\end{figure*}

Crackling noise relation (\cite{sethna2001crackling,jensen1998self}) is used in neuroscience and material science (\cite{mallinson2019avalanches}), for a criterion of whether a system is critical. 
When a system is critical, any physical quantities follow power-laws so that any two critical exponents satisfy the relation (\cite{jensen1998self}).
To examine the relation, we examined the spatial and temporal quantities, distance and duration of walking bouts (Fig.\ref{fig_scaling_relation}).
The bouts were defined as time series of speed above the threshold of 0.25 mm/s. 
If the speed was less than the value, it was assumed that the snails were not moving.
Although the region in which the criterion was satisfied is limited (Fig.\ref{fig_scaling_relation}), the relation was well satisfied.


\section{Discussion}
The results not only showed that the movement patterns of the snails exhibited typical L\'{e}vy walks but also multiple power-laws, and fulfilled the crackling noise relation (\cite{jensen1998self,sethna2001crackling}), a condition of criticality in the brain criticality hypothesis. 
Although the range in which this relation holds is limited, the fact that multiple spatiotemporal quantities follow power laws strongly suggests that the behavior behind these movement patterns is indicative of a critical state of a nonlinear dynamical system. 
We obtained the data from behavioral movements, but the nervous systems, especially the system in soft-bodied organisms, are inseparable from their bodies. 
Even in the study of more harder bodied organism, fluit fly (\cite{maye2007order}), such inseparability has been considered and the dynamics is thought to be generated from the nervous systems.
Furthermore, the observation that this critical state was consistently seen across individual movement patterns, despite the genetic, developmental, and physiological differences, suggests that this critical state is self-organized. 
Additionally, the presence of crabs did not affect this critical state. 
However, further research with more controlled conditions is necessary to understand how this self-organization occurs.

The L\'{e}vy walk is a mathematical model for generating movement patterns determined by probabilistic values from a power-law distribution. 
This model has been reported in various species, including invertebrates species (\cite{reynolds2018current}). 
However, the dynamic mechanism by which these patterns are generated remained unknown. 
There is a proposal that these patterns originate from nonlinear dynamical systems in a critical state (\cite{abe2020functional}).
In this context, data from \textit{Drosophila} larvae indicated that the functioning of the nervous system in a critical state is key. 
Additionally, another study reported that even when the brain function of \textit{Drosophila} was suppressed, leaving only the thoracic central pattern generator, L\'{e}vy walks were still generated (\cite{sims2019optimal}). 
This suggests that these walks are spontaneously generated in the thoracic ganglia. 
Moreover, research using a sophisticated flight simulator has shown that flight patterns in adult \textit{Drosophila}, which follow power laws, are intrinsically generated (\cite{maye2007order}). 
These results strongly suggest that the movement patterns of L\'{e}vy walks are self-organized within the nervous system rather than being influenced by the external environment.

In the nervous systems, particularly in the brains, spontaneous neural activities such as neural avalanches (\cite{beggs2003neuronal,plenz2021self}) and readiness potentials (\cite{schurger2021readiness}) are well-known. 
Both have been recorded in a wide range of animals ``from crayfish to human'' (\cite{schurger2012accumulator,kagaya2010readiness,kagaya2011sequential,kagaya2022self}) and are independent of the recording methods used. 
They can be observed through EEG, extracellular recordings, and intracellular recordings. 
It is considered that both phenomena are manifestations of self-organized critical states (\cite{kagaya2022self}). 
In this study, we presented results that support the brain criticality hypothesis in the snails. 
Although the data we dealt with were behavioral, not neural data,  the inseparability of the dynamic coupling of the brain, body, and environment is likely to allow us to study the neural dynamics from the behavior (\cite{pfeifer2007self}).

\section{Conclusion}
In the future, the establishment of experimental systems that enable simultaneous recording of both neural and behavioral data across various animals is expected to elucidate the neural mechanisms underlying spontaneous behavior and the sensory-motor information processing that occurs within it, from a comparative physiological perspective (\cite{murakami2023toward}). 
This will reveal the unique properties inherent to each species and individual, as well as the properties that are universally present across different species.
Snail-like ``traveling-wave-type'' robot has been developped (\cite{ogawa2012path}). 
Such universality underlying the real organisms can provide an insight into how to implement spontaneity into soft robots.

\section*{Acknowledgments}
We would like to express our gratitude to Mr. Keita Harada for his technical support, including the construction of the container used in the experiments.

\bibliographystyle{plainnat}
\bibliography{ref}

\pagebreak



\end{document}